\documentclass[%
 reprint,
 superscriptaddress,
 amsmath,amssymb,
 aps,showpacs,prl 
twocols,
]{revtex4-1}

\usepackage{graphicx}% Include figure files
\usepackage{dcolumn}% Align table columns on decimal point
\usepackage{bm}% bold math
\usepackage{color}

\begin{document}

\title{Monte Carlo adaptive resolution simulation of multicomponent molecular liquids}

\author{Raffaello Potestio}
\email{potestio@mpip-mainz.mpg.de}
\affiliation{Max-Planck-Institut f\"ur Polymerforschung, Ackermannweg 10, 55128 Mainz, Germany}
\author{Pep Espa\~{n}ol}
\affiliation{Dep.to de F\'{\i}sica Fundamental, Facultad de Ciencias (U.N.E.D.),  Avda. Senda del Rey 9, 28040 Madrid, Spain}
\author{Rafael Delgado-Buscalioni}
\affiliation{Dep.to de F\'{\i}sica Teorica de la Materia Condensada and IFIMAC, Universidad Aut\'{o}noma de Madrid, Campus de Cantoblanco, 28049 Madrid, Spain}
\author{Ralf Everaers}
\affiliation{Laboratoire de Physique et Centre Blaise Pascal, {\'E}cole Normale Sup\'erieure de Lyon, CNRS UMR5672, 46 all\'{e}e d'Italie, 69364 Lyon, France}
\author{Kurt Kremer}
\affiliation{Max-Planck-Institut f\"ur Polymerforschung, Ackermannweg 10, 55128 Mainz, Germany}
\author{Davide Donadio}
\affiliation{Max-Planck-Institut f\"ur Polymerforschung, Ackermannweg 10, 55128 Mainz, Germany}

\begin{abstract}
Complex soft matter systems can be efficiently studied with the help of adaptive resolution simulation methods, concurrently employing two levels of resolution in different regions of the simulation domain. The non-matching properties of high- and low-resolution models, however, lead to thermodynamic imbalances between the system's subdomains. Such inhomogeneities can be healed by appropriate compensation forces, whose calculation requires nontrivial iterative procedures. In this work we employ the recently developed Hamiltonian Adaptive Resolution Simulation method to perform Monte Carlo simulations of a binary mixture, and propose an efficient scheme, based on Kirkwood Thermodynamic Integration, to regulate the thermodynamic balance of multi-component systems.
\end{abstract}

\maketitle

Soft matter systems often display an inherently multiscale nature. Because of this interplay of length and time scales, a unique level of description is not sufficient: a fully atomistic (AT) simulation would be too computationally expensive, while coarse-grained (CG) models \cite{kremer2000,peter_kremer_softmatter,vdvegt2009,Hijon2010,noid_chapter} would lack the necessary detail to account for local interactions.
In recent years methods have been developed that couple models with different resolutions in a single simulation, where a small ``important" region is treated at full atomistic level, while in the surrounding a coarser model is used. Examples of successful applications of this approach are mixed quantum mechanics/molecular mechanics schemes \cite{WARSHEL:1976p4538,qmmm1,Svensson:1996p4537,Carloni:2002p4461,bulo}, also employed to study crack propagation in hard matter \cite{rudd,Rottler:2002,Csanyi:2004,Jiang:2004,kax}, and the extension to complex fluids \cite{heyden:2008,heyden:2009,jcp,ensing,prlcomment2011,adresstf,FritschPRL,adolfoprl,potestio,debashish2,hadress,luigiPRX}, where diffusion plays a crucial role.
Adaptive resolution methods unite the advantageous simplicity and general versatility of CG models with the chemical specificity of higher resolution AT descriptions. To employ them successfully one needs to correct  the thermodynamic mismatch that usually exists between models at different resolution. In the Adaptive Resolution Simulation (AdResS) scheme \cite{jcp,annurev,ensing,prlcomment2011,adolfoprl} this is achieved with the help of a thermodynamic force \cite{adresstf,FritschPRL} numerically obtained by an iterative procedure, which can become involved for multi-component systems \cite{debashish2}. Here we tackle this problem using the recently developed Hamiltonian AdResS \cite{hadress} (H-AdResS) and its tight connection to Kirkwood Thermodynamic Integration (TI) \cite{kirkwood1935}. Our goal is twofold: first, we demonstrate the possibility to perform Monte Carlo (MC) simulations of a nontrivial double-resolution system using H-AdResS, formerly established in the framework of Molecular Dynamics (MD) \cite{hadress}. Second, we describe a strategy to obtain in a single computationally efficient calculation the potential energy functions required to regulate the thermodynamic balance between AT and CG regions for multi-component mixtures.

\begin{figure}
 \includegraphics[width=7.5cm]{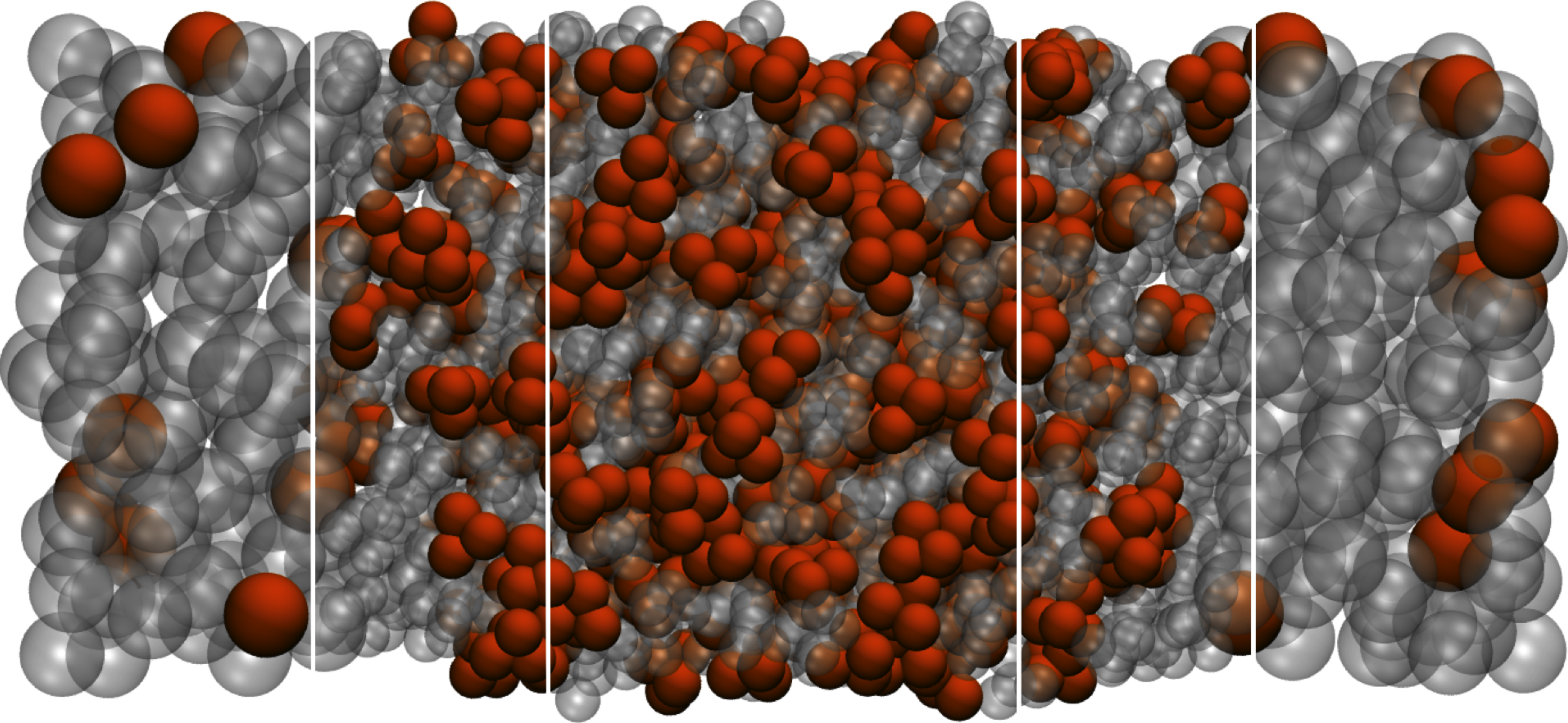}\\
 \includegraphics[width=7.5cm]{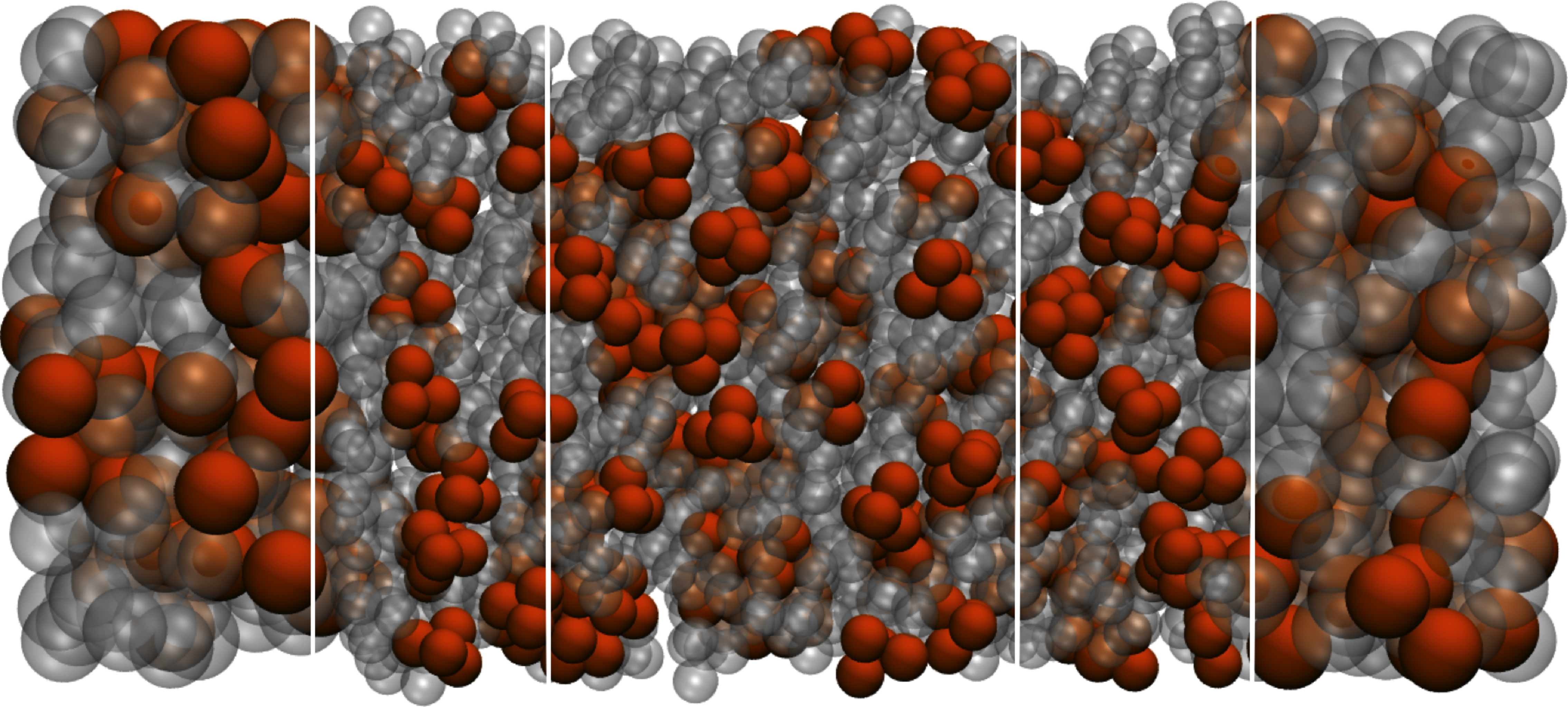}
 \caption{Snapshots of a simulation of Case II. Top panel: equilibrated configuration, without FEC. Bottom panel: equilibrated configuration, with FEC. The A-type atoms are represented in gray, the B-type atoms in orange. Molecules in the coarse-grained (CG) region are represented as large spheres. White vertical lines mark the boundaries of the CG-hybrid and hybrid-atomistic regions.\label{fig:snap_1}}
 \end{figure}

%%%%%%%%%%%%%%%%%%%%%%%%%%%%%%%%%%%%
H-AdResS \cite{hadress} is formulated in terms of a global Hamiltonian $H$, similar in spirit to the one used in TI \cite{kirkwood1935}. In the H-AdResS Hamiltonian the total intermolecular energy of each molecule is weighted with a sigmoid function $\lambda(\textbf{R})$, that depends on the center-of-mass coordinate ${\bf R}$ of the molecule and ranges from 0 (purely CG) to 1 (purely AT):
\begin{equation}\label{eqn:hadress_H}
\begin{split}
&H = K + V^{int} + \sum_{a} \left[\lambda_a {V^{AT}_a} + {(1 -\lambda_a}) {V^{CG}_a} \right]\\
&V_a^{AT} = \frac{1}{2} \sum_{a' \neq a} \sum_{ij} V^{AT}_{a i; a' j}\ ,\ V_a^{CG} = \frac{1}{2} \sum_{a' \neq a} V^{CG}_{aa'}
\end{split}
\end{equation}
\noindent where $i,j$ are atom indices, $\lambda_a = \lambda(\textbf{R}_a)$, $K$ is the all-atom kinetic energy and $V^{int}$ is the intramolecular interaction.
If the Hamiltonian in Eq. \ref{eqn:hadress_H} is straightforwardly used in a simulation, the difference in chemical potential between the AT and CG resolutions determines a density and pressure imbalance between the two regions of the system. In order to restore a flat density profile we introduced a compensation term $\Delta H(\lambda)$ in the Hamiltonian, which then reads $H_{\Delta} = H - \sum_{a} \Delta H(\lambda(\textbf{R}_a))$. $\Delta H(\lambda)$ can be approximated with the Gibbs free energy difference per molecule (chemical potential) $\Delta G/N$, as obtained from a TI of a homogeneous system, performed in the canonical ensemble~\cite{hadress}:
\begin{equation}\label{eqn:fec_2}
\begin{split}
&\Delta H(\lambda) \equiv \frac{\Delta G(\lambda)}{N} = \frac{\Delta F(\lambda)}{N} + \frac{\Delta p(\lambda)}{\rho^\star}\\
&\Delta F(\lambda) = \int_0^{\lambda} d\lambda' \left\langle \left[ V^{AT} - V^{CG} \right]  \right\rangle_{\lambda'}
\end{split}
\end{equation}
where $\rho^\star \equiv N/V$ is the reference molecular number density and $\Delta p(\lambda) = p(\lambda) - p(0)$ is the pressure difference.
The Free Energy Compensation (FEC) strategy, defined by Eq. \ref{eqn:fec_2}, can be extended to multi-component systems. To illustrate this idea we consider a molecular liquid composed by two types of molecules, $A$ and $B$, indexed with $a$ and $b$, respectively. The corresponding H-AdResS Hamiltonian for this system reads:
\begin{equation}\label{eqn:Hmixture}
\begin{split}
H^{MIX} = K + V^{int} &+ \sum_{a \in A} \left [\lambda_a V^{AT}_{a} + (1 - \lambda_a) V^{CG}_{a} \right ]\\
&+ \sum_{b \in B} \left [\lambda_b V^{AT}_{b} + (1 - \lambda_b) V^{CG}_{b} \right ]
\end{split}
\end{equation}
with $\lambda_a = \lambda(\textbf{R}_a)$ and $\lambda_b = \lambda(\textbf{R}_b)$. The intermolecular potential energy terms are given by the following expressions:
\begin{equation}\label{eqn:Hmixture2}
\begin{split}
V^{AT}_{a} &= \frac{1}{2} \left [ \sum_{ \substack{a' \in A \\ a' \neq a}} \sum_{ij} V[AA]^{AT}_{a i; a' j} + \sum_{b \in B} \sum_{ij} V[AB]^{AT}_{a i; b j} \right ]\\ 
V^{CG}_{a} &= \frac{1}{2} \left [ \sum_{ \substack{a' \in A \\ a' \neq a}} V[AA]^{CG}_{aa'} + \sum_{b \in B} V[AB]^{CG}_{ab} \right ]\\
V^{AT}_{b} &= \frac{1}{2} \left [ \sum_{ \substack{b' \in B \\ b' \neq b}} \sum_{ij} V[BB]^{AT}_{b i; b' j} + \sum_{a \in A} \sum_{ij} V[AB]^{AT}_{b i; a j} \right ]\\ 
V^{CG}_{b} &= \frac{1}{2} \left [ \sum_{ \substack{b' \in B \\ b' \neq b}} V[BB]^{CG}_{bb'} + \sum_{a \in A} V[AB]^{CG}_{ba} \right ]
\end{split}
\end{equation}
\noindent where $V[XY]$ is the non-bonded interaction between a molecule of type $X$ and a molecule of type $Y$, with $X,Y = A, B$, and the indices $i,j$ label the atoms.

In analogy with one-component systems we introduce a FEC term for each species to compensate for the free energy difference between the AT and the CG regions:
\begin{equation}\label{eqn:delta_Hmixture}
H^{MIX}_{\Delta} = H^{MIX} - \sum_{a \in A} \Delta H_A(\lambda_a) - \sum_{b \in B} \Delta H_B(\lambda_b)
\end{equation}
An {\it Ansatz} for the compensation term of a given species $k =a, b$ can be obtained from TI as follows:
\begin{equation}\label{eqn:fec_mix}
\begin{split}
&\Delta H_k(\lambda)= \frac{\Delta F_k(\lambda)}{N_k} + \frac{\Delta p_k(\lambda)}{\rho_k^\star}\\ 
&\Delta F_k(\lambda) = \int_0^\lambda d\lambda' \left\langle \left[ V^{AT}_k - V^{CG}_k \right] \right\rangle_{\lambda'}\\ 
&\Delta p_k(\lambda) = p_k(\lambda) - p_k(0)
\end{split}
\end{equation}
\noindent where the $N_k$, $\rho_k^\star \equiv N_k/V$ and $p_k$ are, respectively, the number of molecules, the reference partial density and the partial virial pressure of species $k$ \cite{SI}. We stress that all the quantities in Eq. \ref{eqn:fec_mix} can be computed in a single TI of the mixture from AT to CG at the concentration of interest, irrespective of the number of species. All the cross-interactions between different types of molecules are automatically included in the free energy contribution of each species  (details in the Supplemental Information \cite{SI}). Additionally, the Free Energy Compensation $\Delta H_k(\lambda)$ is an {\it intensive} quantity and does not depend on the specific geometry of the H-AdResS setup. It is therefore possible to perform the TI in a relatively small system, provided that it is statistically representative, {\it i.e.} finite size effects are negligible.

%%%%%%%%%%%%%%%%%%%%%%%%%%%%%%%%%%%%
A MC adaptive resolution simulation approach was formerly developed \cite{abrams_adaptive_MC} by introducing a ``dual-resolution partition function'', in which the resolution of a given molecule is a stochastic variable that depends on its position in space.  Although physically sound, this approach cannot be rephrased in terms of a general Hamiltonian, as it is based on a modification of the partition function. In contrast H-AdResS is based on a dual-resolution Hamiltonian, and it can therefore be generalized to any statistical ensemble and simulation technique, thus including MC.
We implemented H-AdResS in a code based on the MC Metropolis algorithm~\cite{metropolis1953}. Our code was tested on the homogeneous fluid used in Ref. \cite{hadress} (data not shown), then used to validate the FEC method for mixtures, following Eqs.~\ref{eqn:delta_Hmixture} and~\ref{eqn:fec_mix}.
Specifically we considered two cases of binary mixtures of tetrahedral molecules, both made of four identical atoms (one of species A and one of species B), connected by quartic anharmonic bonds \cite{jcp,SI}. In Case I the two molecular species are present in equal proportions ($399$ molecules of each type); atoms of the same species interact with a purely repulsive WCA potential, but the effective size of the B-type is larger than the A-type. In Case II, $70\%$ ($558$) A-type molecules and $30\%$ ($240$) B-type molecules were used. In contrast with Case I, the A--A and B--B WCA interactions are identical. In both cases the A--B interaction is a Lennard-Jones potential. The simulations were performed in the $NVT$ ensemble at a temperature $T = 120 K$. The dimensions of the simulation box are $L_x = 3.684\ nm$, $L_y = L_z = 1.50\ nm$, with periodic boundary conditions in all directions. The atomistic/hybrid interface and the hybrid/coarse-grained interfaces are located at $d_h = \pm 0.15 L_x$ and $d_h + s_h = \pm 0.3 L_x$ from the box center, respectively. More details about the simulation setup are provided in \cite{SI}.

\begin{figure}
 \includegraphics[width=8cm]{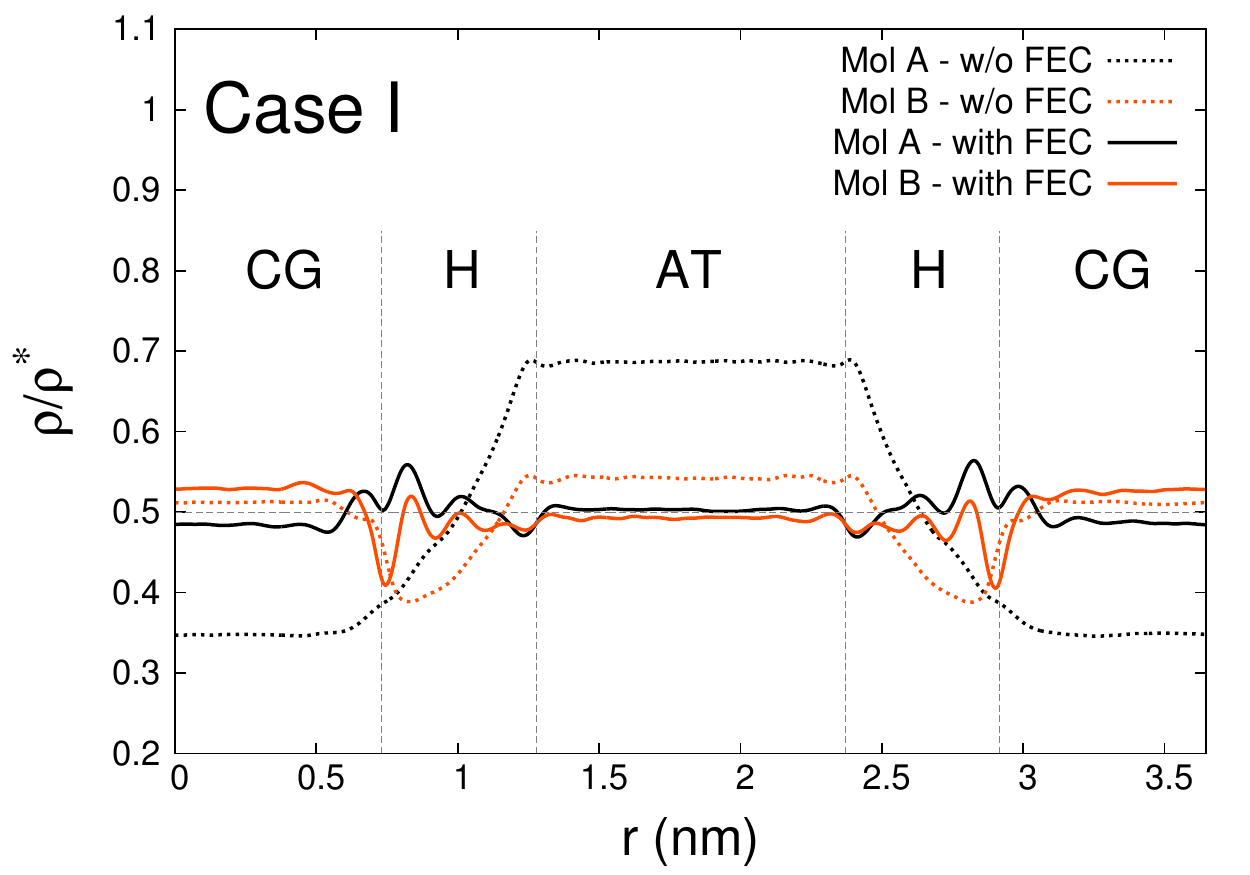}\\
 \includegraphics[width=8cm]{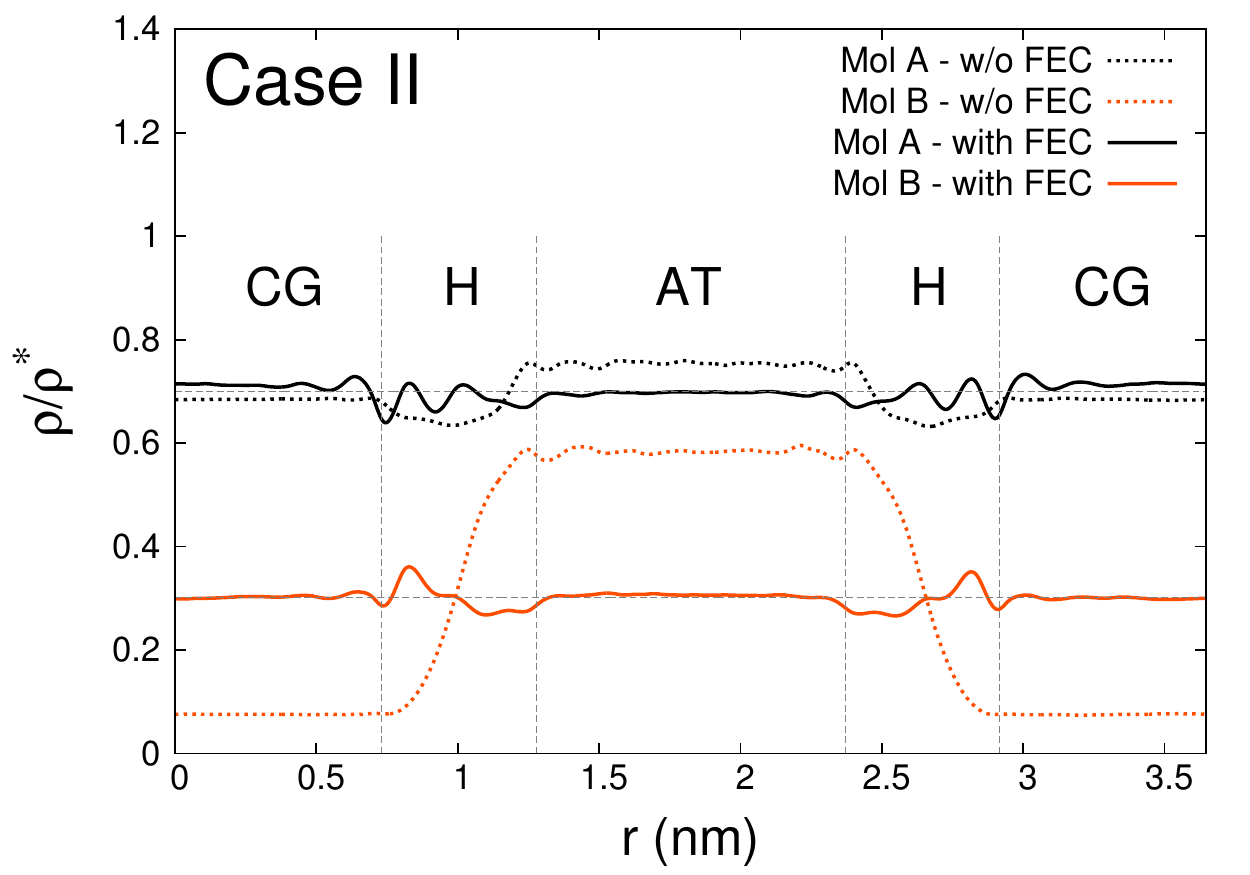}
 \caption{Density profiles along the direction of resolution change, for Case I (top) and Case II (bottom). Dotted lines: H-AdResS simulations without FEC; solid lines: with FEC. Vertical dashed lines indicate the boundaries between the AT, hybrid and CG regions; horizontal dashed lines mark the reference value of the density (normalized to the total density) as expected in a fully atomistic simulation of the system.\label{fig:dens}}
 \end{figure}

As for the CG model one could choose among a number of different strategies \cite{kremer2000,peter_kremer_softmatter,vdvegt2009,Hijon2010,noid_chapter}, each of which targets a specific property of the underlying AT system. %Here we did not seek for CG interaction potentials that faithfully reproduce some properties of the atomistic system. 
Here instead our intent is to show that H-AdResS and the FEC method allow a completely general and flexible coupling, irrespective of the specific CG potential used. To this end we use the {\em same} CG model in both cases, representing molecules as spherical particles with identical, purely repulsive WCA A--A, B--B and A--B interactions \cite{SI}. The resulting thermodynamic mismatch in chemical potentials between AT and CG domains is particularly large in Case II: simple visual inspection (Fig. \ref{fig:snap_1} top) is in fact sufficient to detect a large accumulation of B-molecules in the AT zone. Closer inspection of the density profiles (dotted lines in Fig. \ref{fig:dens}) also reveals significant deviations in Case I. As a consequence, neither the total density nor the relative concentrations in the AT zone obtained using the uncompensated adaptive resolution Hamiltonian in Eq.~\ref{eqn:Hmixture} correspond to the reference atomistic system.

\begin{figure}
 \includegraphics[width=8cm]{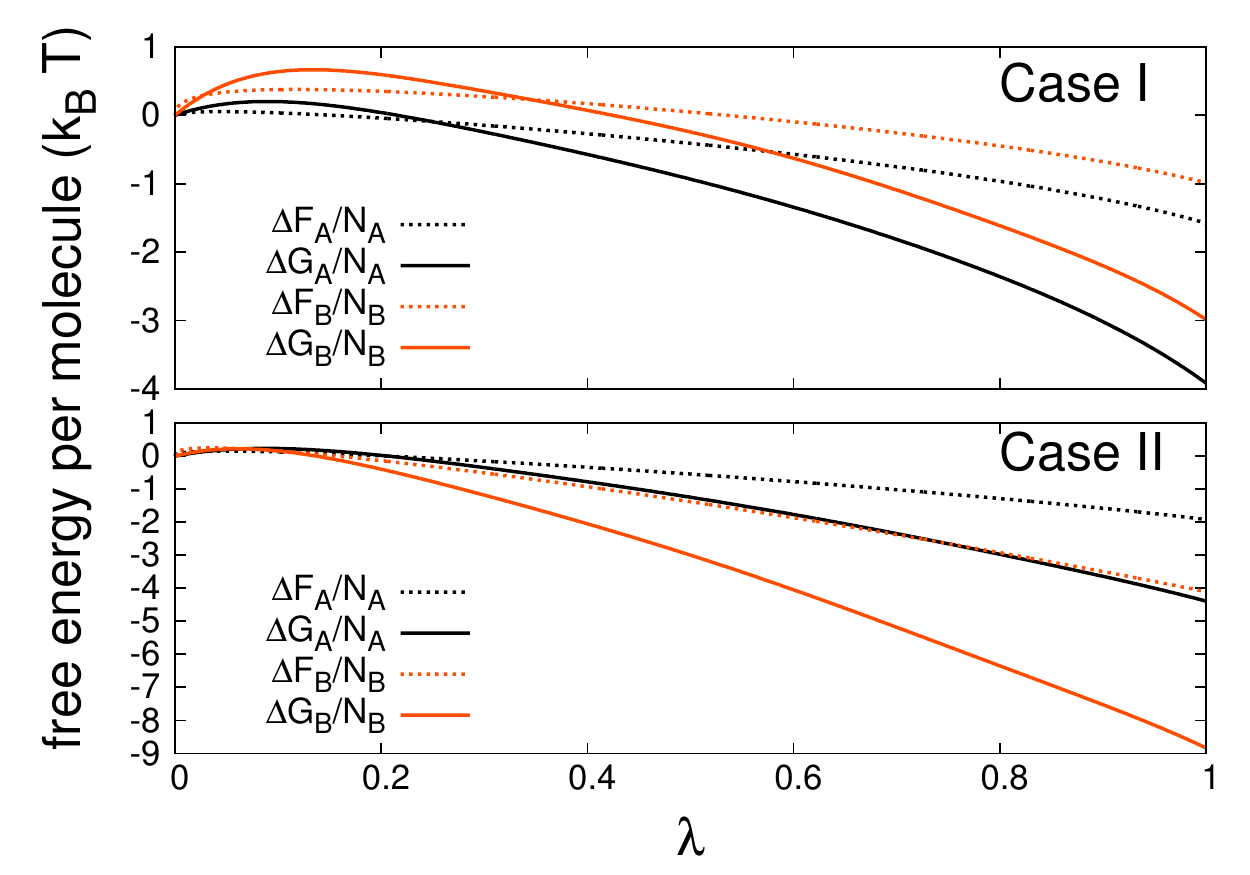}
 \caption{Free energy differences per molecule between the AT and CG models as a function of the mixing parameter $\lambda$ for Case I (top) and Case II (bottom). The Helmholtz free energy is represented by the dotted lines, the Gibbs free energy by the solid lines. Molecular species A corresponds to the black curves, the species B to the orange curves.
 \label{fig:fecs}}
 \end{figure}

According to Eq. \ref{eqn:fec_mix}, we have determined the thermodynamic mismatch between the AT and the CG zone from two TI runs (one per case studied) where we switch the interactions of the mixtures from purely CG ($\lambda=0$) to purely AT ($ \lambda=1$). The Helmholtz and Gibbs free energy differences per molecule between the CG and AT models as a function of the coupling parameter $\lambda$, computed for both species {\it simultaneously} in a single TI for each case, are shown in Fig. \ref{fig:fecs}. In Case I both Helmholtz and Gibbs free energy differences are similar in shape for the two species. The B-type molecules show a Gibbs free energy difference per particle $\Delta G_B/N_B \equiv (G_B(1) - G_B(0))/N_B$ smaller in magnitude by $\sim 1\ k_B T$ than the A-type molecules. This difference can be attributed to the larger effective size of B-type atoms compared to A-type ones, which makes the size of the B-type AT molecules closer to that of the CG model. In Case II the situation is remarkably different. In spite of the same interaction between molecules of the same type ($V[AA] \equiv V[BB]$), the uneven relative concentration of the two species determines a much larger free energy difference between the AT and CG models for the B-type. In fact, the latter shows a Gibbs free energy difference per particle $|\Delta G_B/N_B| > 2\ |\Delta G_A/N_A|$. This is mainly due to the fact that the interaction between A and B types is attractive only in the AT representation, thus determining a lower chemical potential for the minority type (B) in the AT region. In addition, in both cases the sign of $\Delta G$ favors the densification of particles in the AT region, as it can be seen in Fig. \ref{fig:dens}.

To counterbalance the mismatch in chemical potentials we introduce a FEC in the H-AdResS Hamiltonian according to Eq. \ref{eqn:delta_Hmixture}, using the free energy functions shown in Fig. \ref{fig:fecs}. The resulting density profiles (solid lines in Fig. \ref{fig:dens}) demonstrate the success of the procedure: in both Case I and II the densities of the two species attain, in the AT region, the same values that would be observed in a fully-atomistic simulation (for Case II see also Fig. \ref{fig:snap_1} bottom); also the pairwise correlation functions in the AT region of both cases perfectly superimpose to the all-atom reference (see SI~\cite{SI}). In particular, it is remarkable that the simple FEC strategy fixed the density of B-type molecules in Case II, where the free energy difference per particle between the AT and the CG representations span over one order of magnitude. In the CG region of Case I a small ($\sim3 \%$) deviation from the reference can be observed, due to the depletion in the hybrid region typical of adaptive resolution simulations \cite{jcp,adresstf,FritschPRL}. These density fluctuations are due to correlations between close-by molecules at different resolutions, which the FEC method, based on TI simulation where $\lambda$ is the same for all molecules, cannot capture. These ripples, however, affect only the hybrid region (see Fig. \ref{fig:dens}), and can be leveled out employing iterative methods \cite{FritschPRL} to correct the FEC functions.
\begin{figure}
 \includegraphics[width=8cm]{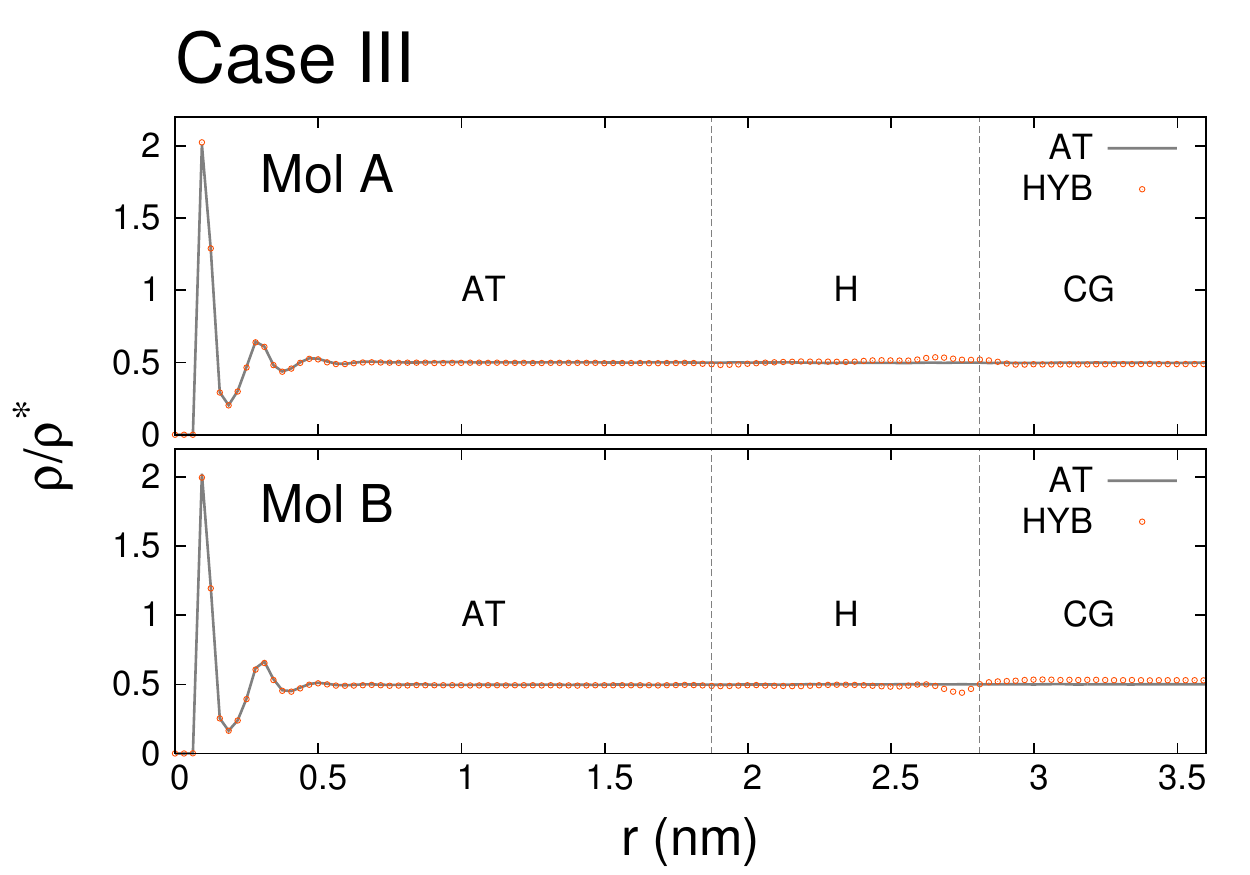}
 \caption{Density profiles of the system of Case III, where the mixture is put in contact with an attractive, fixed and non-permeable wall. The results obtained in a H-AdResS simulation (orange points) perfectly superimpose on those of an analogous all-atom simulation (gray lines) for both A and B molecules (top and bottom panel, respectively).
 \label{fig:wall}}
 \end{figure}
A further validation of the effectiveness of the FEC method is provided in Case III, where we performed the simulation of a system analogous to that of Case I put in contact with a fixed, attractive wall. 
The latter is implemented as a Lennard-Jones potential acting in the same way on all the molecules, and depends on the distance between each atom and the wall. This potential has the same $\sigma$ as the A--B interaction, but is four times stronger.
In this case, only the subregion of the system close to the wall is treated at the atomistic level, and the same FEC's used for the homogeneous system were employed (details in SI~\cite{SI}). In Fig. \ref{fig:wall} we report the results obtained from this simulation: for both molecular species the density profiles superimpose perfectly on the reference, calculated from a fully-atomistic simulation.

Finally we mention two technical but relevant aspects of the dual-resolution approach, namely the sampling efficiency and the computational speedup. Using a CG model with softer interactions compared to the AT model we enhance the acceptance rate of the MC moves, thus improving the sampling of the configurational space. Additionally, the reduction of the number of degrees of freedom and the usage of simple CG potentials reduces the CPU time by a factor proportional to the level of coarse-graining and to the system size $L$. For example, when a slab geometry is employed the computational gain grows linearly with $L$, while using a spherical AT region the speedup grows as $L^3$  \cite{SI}.

%%%%%%%%%%%%%%%%%%%%%%%%%%%%%%%%%%%%
In summary, we employed the H-AdResS scheme to perform dual-resolution Monte Carlo simulations, and extended the FEC technique based on TI to regulate the density balance of AT and CG subregions, to the general case of a multi-component system. This method allowed us to couple a two-component AT system to a CG potential, in which the two species are indistinguishable. In spite of the large free energy difference existing between these two models, the FEC approach effectively compensates large density imbalances. 
%In contrast to iterative force-based approaches \cite{debashish2}, 
This procedure seamlessly accounts for the correlations between the densities of the various species. This work thus lays the theoretical background to drastically simplify the steps required to perform dual-resolution simulations of multicomponent systems, coupling an atomistic complex fluid to a simple CG model whose thermodynamic properties do not match the atomistic reference, 
still preserving the reference thermodynamic properties in the AT region. The limited impact of the choice of the CG potential 
%on the AT region 
can therefore be exploited to our advantage, for example by choosing CG potentials that facilitate either grand canonical particle insertion or particle switch in semi-grand-canonical simulations.
Possible applications range from crystal growth \cite{urea1,urea2} in an effective grand canonical ensemble to free energy calculations of biological systems in aqueous solutions \cite{debashish2}. H-AdResE also offers a promising route to fast calculations of variations of solvation free energy differences ($\Delta \Delta G$) due to changes in the composition and/or structure of a solvated macromolecule. 
%At the same time, in the present work we have also validated the H-AdResS method, previously successfully applied in the framework of Molecular Dynamics, to Monte Carlo simulation algorithms. 
In addition, the validation of H-AdResS in the framework of MC is of particular relevance for hybrid quantum/classical simulations \cite{adolfoprl,potestio} based on Path Integrals \cite{feynman2,tuckermann_book}, where the inherently energy-based formulation makes it natural to employ MC.

%%%%%%%%%%%%%%%%%%%%%%%%%%%%%%%%%%%%
\begin{acknowledgments}
The authors thank K. Kreis for a critical reading of the manuscript, and acknowledge hospitality at KITP, where this collaboration was initiated. This research was supported in part by the National Science Foundation under Grant No. NSF PHY11-25915. PE thanks the support  of BIFI and the  Ministry of Science and Innovation  through project  FIS2010-22047-C05-03. RDB also thanks FIS2010-22047-C05-01 and the support  of the ``Comunidad de Madrid'' via the project MODELICO-CM (S2009/ESP-1691).
\end{acknowledgments}
%%%%%%%%%%%%%%%%%%%%%%%%%%%%%%%%%%%%
\bibliography{bibliography}

\end{document}